\documentclass[sigplan,screen]{acmart}

\usepackage[]{hyperref}
\usepackage[]{hyperxmp}

\usepackage[most]{tcolorbox}
\usepackage{multirow}
\usepackage{makecell}
\usepackage{bm}
\usepackage{subfig}

\AtBeginDocument{%
  }

\copyrightyear{2025}
\acmYear{2025}
\setcopyright{acmlicensed}\acmConference[ASPLOS '25]{Proceedings of the 30th ACM International Conference on Architectural Support for Programming Languages and Operating Systems, Volume 3}{March 30-April 3, 2025}{Rotterdam, Netherlands}
\acmBooktitle{Proceedings of the 30th ACM International Conference on Architectural Support for Programming Languages and Operating Systems, Volume 3 (ASPLOS '25), March 30-April 3, 2025, Rotterdam, Netherlands}
\acmDOI{10.1145/3676642.3736117}
\acmISBN{979-8-4007-1080-3/2025/03}

\begin{document}

\title[ASDR: Exploiting \underline{A}daptive \underline{S}ampling and \underline{D}ata \underline{R}euse for CIM-based\\ Instant Neural Rendering]{ASDR: Exploiting \underline{A}daptive \underline{S}ampling and \underline{D}ata \underline{R}euse for CIM-based Instant Neural Rendering}

\author{Fangxin Liu}
\authornote{Both authors contributed equally to the paper.}
\authornote{Corresponding authors.}
\orcid{0000-0002-8769-293X}
\affiliation{
    \institution{Shanghai Jiao Tong University}
    \institution{Shanghai Qi Zhi Institute}
    \city{Shanghai}
    \country{China}
}
\email{liufangxin@sjtu.edu.cn}

\author{Haomin Li}
\authornotemark[1]
\orcid{0000-0002-2939-6534}
\affiliation{
    \institution{Shanghai Jiao Tong University}
    \institution{Shanghai Qi Zhi Institute}
    \city{Shanghai}
    \country{China}
}
\email{haominli@sjtu.edu.cn}

\author{Bowen Zhu}
\orcid{0009-0003-2957-469X}
% \affiliation{
%     \institution{Shanghai Jiao Tong University}
%     \city{Shanghai}
%     \country{China}
% }

\author{Zongwu Wang}
\orcid{0009-0003-2157-4927}
\affiliation{
    \institution{Shanghai Jiao Tong University}
    \city{Shanghai}
    \country{China}
}
\email{zhubowen@sjtu.edu.cn}
\email{wangzongwu@sjtu.edu.cn}

\author{Zhuoran Song}
\orcid{0000-0002-6494-4786}
\affiliation{
    \institution{Shanghai Jiao Tong University}
    \city{Shanghai}
    \country{China}
}
\email{songzhuoran@sjtu.edu.cn}

\author{Haibing Guan}
\orcid{0000-0002-4714-7400}
\affiliation{
    \institution{Shanghai Jiao Tong University}
    \city{Shanghai}
    \country{China}
}
\email{hbguan@sjtu.edu.cn}

\author{Li Jiang}
\authornotemark[2]
\orcid{0000-0002-7353-8798}
\affiliation{
    \institution{Shanghai Jiao Tong University}
    \institution{Shanghai Qi Zhi Institute}
    \city{Shanghai}
    \country{China}
}
\email{ljiang_cs@sjtu.edu.cn}

% \renewcommand{\shortauthors}{Fangxin Liu, Haomin Li, Bowen Zhu, Zongwu Wang, et al.}

\renewcommand{\shortauthors}{Fangxin Liu et al.}
%% No italics, no superscripts
%% Use footnote or author note to identify equal contribution and/or contact author info

\begin{abstract}
Neural Radiance Fields (NeRF) offer significant promise for generating photorealistic images and videos. However, existing mainstream neural rendering models often fall short in meeting the demands for immediacy and power efficiency in practical applications. Specifically, these models frequently exhibit irregular access patterns and substantial computational overhead, leading to undesirable inference latency and high power consumption. Computing-in-memory (CIM), an emerging computational paradigm, has the potential to address these access bottlenecks and reduce the power consumption associated with model execution.

To bridge the gap between model performance and real-world scene requirements, we propose an algorithm-architecture co-design approach, abbreviated as ASDR, a CIM-based accelerator supporting efficient neural rendering. 
At the algorithmic level, we propose two rendering optimization schemes: 
(1) Dynamic sampling by online sensing of the rendering difficulty of different pixels, thus reducing access memory and computational overhead. 
(2) Reducing MLP overhead by decoupling and approximating the volume rendering of color and density. 
At the architecture level, we design an efficient ReRAM-based CIM architecture with efficient data mapping and reuse microarchitecture.
Experiments demonstrate that our design can achieve up to $9.55\times$ and $69.75\times$ speedup over state-of-the-art NeRF accelerators and Xavier NX GPU in graphics rendering tasks with only $0.1$ PSNR loss.
\end{abstract}





\maketitle

\section{Introduction}

Neural Radiance Fields (NeRF)~\cite{nerf} represent a significant advancement in bridging the gap between computer graphics and vision by leveraging deep neural networks (DNNs) to reconstruct photo-realistic images and videos with remarkable fidelity. Unlike traditional methods that rely on 3D grids and voxels, NeRF utilizes the powerful generalization capabilities of neural networks to implicitly reconstruct 3D scenes from a limited set of 2D images. This allows for high-quality 2D rendering from novel perspectives, making NeRF particularly useful in applications such as image super-resolution, novel view synthesis, and 3D reconstruction~\cite{wang2022nerf, tancik2022block, 10.1145/3528233.3530718}.

NeRF achieves this by embedding scene and object information within the weights of the neural network, effectively mapping input coordinates to colors and radiance. This process enables NeRF to capture intricate details of complex surfaces and object shapes. The neural rendering process involves learning the continuous volume density and color distribution of a scene by analyzing 2D images captured from various viewpoints. Unlike traditional methods, neural rendering does not generate a standard 3D model file. Instead, it trains a model capable of predicting RGB values and volume density based on 3D positions and viewing direction parameters. The model employs multi-resolution embedding tables to store voxel features~\cite{instngp, yu2021plenoctrees}, which are used to encode input points. These encoded features are then processed by multilayer perceptrons (MLPs) to predict color and density. During the rendering phase, the model samples numerous 3D points along rays emitted from a viewpoint, with each ray corresponding to a pixel in the final 2D image. These sampled points are input into the model to predict their respective RGB values and densities. The predicted information is then used for volume rendering along the rays, ultimately producing the final pixel colors of the image.

\begin{figure}[tp]
    \setlength{\abovecaptionskip}{1pt}
    \setlength{\belowcaptionskip}{2pt}
    \centering
    \includegraphics[width=1.0\linewidth]{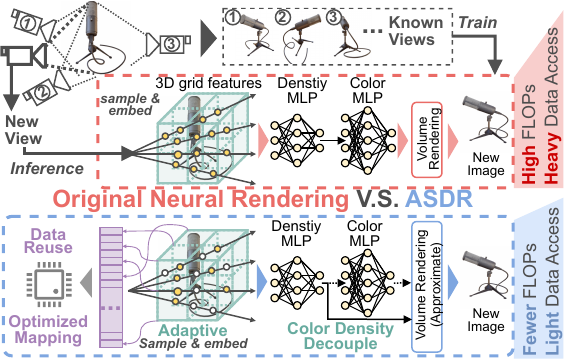}
    \caption{\small Illustration of neural rendering process.
    The \textcolor[RGB]{234, 107, 102}{red} part represents the original neural rendering, while the \textcolor[RGB]{108, 142, 191}{blue} part represents our proposed ASDR rendering process via four hardware-software co-designs: \textcolor[RGB]{103, 171, 159}{adaptive sampling}, \textcolor[RGB]{103, 171, 159}{color density decoupling}, \textcolor[RGB]{166, 128, 184}{data reuse}, and \textcolor[RGB]{166, 128, 184}{mapping optimization}. 
    }
    \label{fig: overview}
    \vspace{-0.45cm}
\end{figure}

Despite the impressive capabilities of NeRF and neural rendering in generating high-quality graphics, they face significant challenges due to their computational and memory demands. The leading NeRF model, such as Instant-NGP~\cite{instngp}, utilizes multi-resolution hash encoding and complex feature computation, requiring access to large hash tables and traversal through MLP networks for every sampled point along each ray. This process imposes substantial demands on memory and computational resources, resulting in low rendering performance even on high-end consumer GPUs. Consequently, optimizing NeRF models for efficiency is crucial for their practical deployment in applications such as virtual reality (VR), augmented reality (AR), and robotics~\cite{li2022rt, 10341922, Zhou_2023_CVPR}, where rendering speed and quality are paramount. 
In VR/AR applications, a frame refresh rate of at least 120Hz is essential to ensure a seamless user experience and prevent dizziness~\cite{wang2023effect}. However, current neural rendering models struggle to meet this requirement. For example, Instant-NGP achieves only ~60 FPS on a high-end GTX 3090 GPU\footnote{\url{https://www.nvidia.com/en-us/geforce/graphics-cards/30-series/}.}, which is half the required frame rate. Additionally, VR/AR devices typically operate under strict power constraints, consuming less than 30W\footnote{\url{https://www.apple.com/apple-vision-pro/specs/}.}, while GPUs like the GTX 3090 consume significantly more power. This substantial gap in both performance and efficiency highlights the challenges in adapting neural rendering for real-world VR/AR applications.

To further enhance the performance of deployed neural rendering models, several prior efforts have been developed. For instance, Inst-3D~\cite{Instant-3D} and Cambricon-R~\cite{Cambricon-R} focus on improving the training efficiency of Instant-NGP. NeuRex~\cite{neurex}, on the other hand, analyzes the inference process of Instant-NGP and implements a subgrid-based method for more hardware-friendly encoding. This technique partitions the input coordinate grid into multiple subgrids, allowing only part of the hash table to be loaded into the on-chip buffer at a time. 

Unfortunately, these techniques end up missing the chance to achieve peak performance: \textbf{C1: }\emph{Overlooking Pixel Difficulty Variation.} Existing methods focus primarily on optimizing the computation for each sampled point, neglecting the differences in rendering difficulty among pixels (i.e., in the number of sample points required by different pixels). \textbf{C2: }\emph{Underutilized Spatial Locality}. The sampled points required for neighboring rays exhibit spatial locality, yet existing methods fail to leverage this data reuse effectively. \textbf{C3: }\emph{Single-Stage Optimization.} Existing methods only focus on a single stage of neural rendering, such as volume rendering or MLP execution, without considering the benefits of co-optimizing across multiple stages. \textbf{C4: }\emph{Irregular \& Heavy Memory Access.} Frequent and irregular lookups in embedding tables lead to significant data handling overhead, impacting overall efficiency. Each of these brings new opportunities for accelerating the rendering (inference) process of NeRF.

\begin{figure*}[tp]
    \setlength{\abovecaptionskip}{1pt}
    \setlength{\belowcaptionskip}{1pt}
    \centering
    \includegraphics[width=0.90\linewidth]{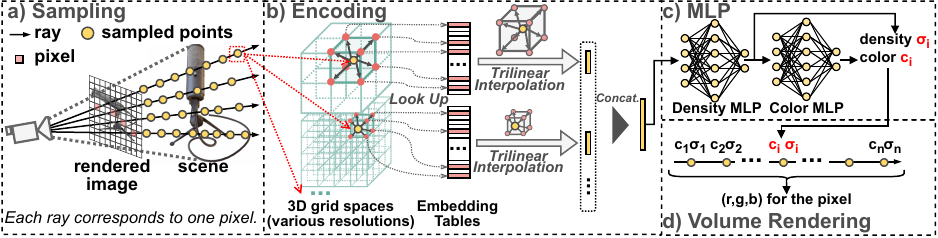}
    \caption{\small Rendering process of Instant-NGP~\cite{instngp}.}
    \label{fig: instngp}
    \vspace{-0.2cm}
\end{figure*}

In this paper, we introduce ASDR, an algorithm-hardware co-design framework to tackle these challenges. ASDR mainly focuses on the optimization of entire stages of neural rendering—encoding, MLP execution, and volume rendering---at both the pixel and sampled point granularities to reduce computation and enhance rendering performance. The key insight behind ASDR is that \emph{leveraging the varying rendering difficulty across pixels and the spatial locality of sampled rays enables us to enhance rendering performance while preserving image quality.} At the algorithm level, as illustrated in Figure~\ref{fig: overview}, we introduce adaptive sampling methods to accommodate the varying rendering difficulty among pixels, adjusting the number of sampling points required for each pixel accordingly (\textbf{C1\&C2}). Additionally, we address the imbalance between the computational demands of color MLP and density MLP by analyzing the volume rendering process and proposing an approximation-based scheme that decouples the execution of density MLP and color MLP (\textbf{C3}). 
At the hardware level, we focus on Computing-in-Memory (CIM)-based architecture, which performs Matrix-Vector Multiplications (MVMs) inside the memory, minimizing the need for extensive data movements of weights and enabling efficient neural network inference. Furthermore, we analyze data access address generation and the distribution of sampled points during encoding. We propose a data reuse and mapping strategy to reduce access conflicts within the CIM-based architecture (\textbf{C4}). Evaluations demonstrate ASDR can deliver a $21.03\times$ speed-up over state-of-the-art NeRF accelerators, with negligible rendering quality loss.
To summarize, we have made the following contributions:
\begin{itemize}
    \item \textbf{Adaptive Sampling Strategy.} We exploit the varying rendering difficulty across pixels by designing a pixel rendering difficulty-aware adaptive sampling strategy. This approach reduces computational requirements while maintaining high rendering quality. (\textbf{C1\&C2})
    \item \textbf{MLP Decoupling Scheme.} We delve into the volume rendering and MLP execution stages, proposing an MLP decoupling scheme based on volume rendering approximation. This strategy significantly reduces the computational overhead of MLP execution. (\textbf{C3})
    \item \textbf{CIM-enabled Data Reuse Architecture.} We propose an efficient CIM-based architecture that addresses the challenges of irregular data access. By integrating an effective data mapping and reuse scheme, we improve resource utilization and alleviate the data access bottleneck, enhancing overall system performance. (\textbf{C4})
\end{itemize}

\section{Background}

\subsection{Neural Rendering Basis}
\label{sec: 2.1}

Neural Rendering is a technique used for modeling a 3D scene based on a set of 2D images with known camera viewpoints. It synthesizes new 2D images based on novel camera viewpoints, a process known as novel view synthesis~\cite{nerf}. As illustrated in Figure~\ref{fig: instngp}, the neural rendering process consists of four main phases: sampling, encoding, color and density prediction, and volume rendering. In the sampling phase, rays are emitted from the camera viewpoint to the imaging plane, with each pixel on the plane corresponding to a specific ray. For each ray, hundreds of points are sampled along its path to capture the scene's details. These sampled points are encoded based on their 3D coordinates to extract spatial features, which are then processed by multilayer perceptrons (MLPs) for further analysis. During the color and density prediction phase, two separate MLPs are used to predict the color and density for each sampled point. The color MLP estimates the RGB values, while the density MLP calculates the volume density for the points. Once the color $c_i$ and densities $\sigma_i$ for all sampled points along a ray are obtained, volume rendering integrates these values to produce the final color $C$ of the corresponding pixel, as shown in Eq.~(\ref{eqn:1}).
\begin{equation}
    C=\sum_{i=1}^{N}T_i\alpha_ic_i \ , \ \ \ \ T_i=\prod\limits_{j=1}^{i-1}(1-\alpha_j)
    \label{eqn:1}
\end{equation}
where $\alpha_i=1-exp(-\sigma_i\delta_i)$ and $\delta_i$ indicates the distance between adjacent sampled points.

\begin{figure*}[tp]
    \setlength{\abovecaptionskip}{1pt}
    \setlength{\belowcaptionskip}{1pt}
    \centering
    \includegraphics[width=0.95\linewidth]{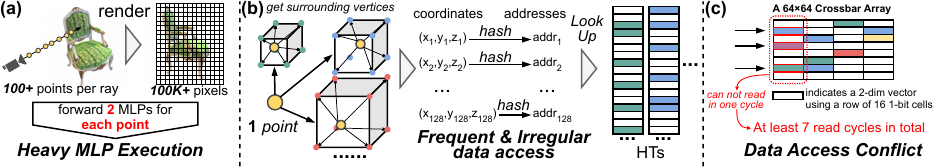}
    \caption{\small Bottleneck Analysis. (a) Heavy MLP Execution: The number of MLP executions is linearly related to the number of sampled points. (b) Frequent and Irregular Data Accesses During Encoding: A sample point is localized into voxels in 16 resolution 3D spaces, with each voxel consisting of 8 vertices requiring embedding lookups. Hashing of coordinates leads to irregular data access. (c) Data Access Conflict on the Hash Table: The hash table can only process one line at a time.}
    \vspace{-0.32cm}
    \label{fig: analysis}
\end{figure*}

\begin{figure}[tp]
    \setlength{\abovecaptionskip}{1pt}
    \setlength{\belowcaptionskip}{1pt}
    \centering
    \includegraphics[width=0.95\linewidth]{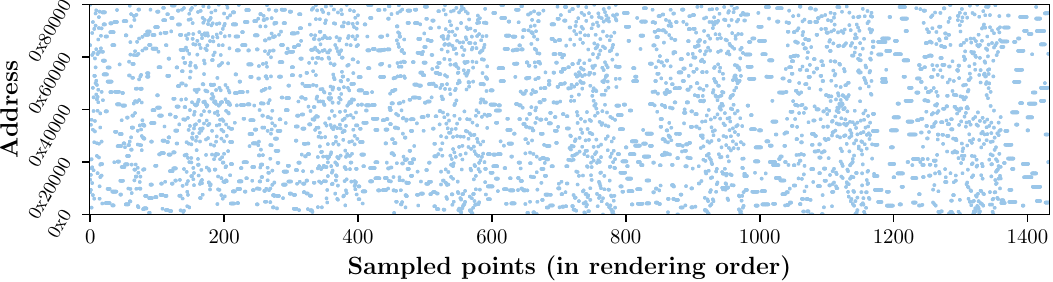}
    \caption{\small Data Access Visualization.}
    \label{fig: access-analysis}
    \vspace{-0.3cm}
\end{figure}

\subsection{Neural Rendering Models}
\label{sec: 2.2}

Representative rendering Models such as NeRF~\cite{nerf}, NeRF-SH~\cite{yu2021plenoctrees}, NSVF~\cite{liu2020neural}, and Instant-NGP~\cite{instngp} can be categorized into two main groups: the original NeRF-based models and parametric encoding-based model.

\textbf{\textit{Original NeRF-based models.}} The use of cosine functions to encode the sampled points in vanilla NeRF does not yield much gain information, resulting in the need for huge MLPs to complete the color and density modeling.

\textbf{\textit{Parametric encoding-based models.}} Instant-NGP, a leading example of parametric encoding-based models, introduces the novel multi-resolution hash encoding, which has become the mainstream NeRF model~\cite{yu2021plenoctrees, instngp,sun2022direct,tensorf}. The key distinction between Instant-NGP and the original NeRF models lies in its use of multi-resolution hash tables for encoding points. This approach allows for a significantly smaller MLP, which reduces computational costs and enhances both training and inference speed, while still maintaining high-quality renderings.
Specifically, each vertex's features in a 3D grid are stored in a hash-based embedding table of size $T$. For a vertex with coordinates $(x,y,z)$, its index in the embedding table is computed described in Eq. (\ref{eqn:2}), where $\oplus$ denotes the bit-wise XOR operation, $\bf{\pi_i}$ are unique, large prime numbers.
\begin{equation}
    index = (x\times \pi_1)\oplus (y\times \pi_2) \oplus (z\times \pi_3)\ mod\ T
    \label{eqn:2}
\end{equation}

For a sample point, Instant-NGP retrieves its corresponding voxel in each resolution space and looks up the features of the eight vertices of the voxel from the embedding table. The features of the sample point in each resolution space are computed using tri-linear interpolation, which leverages the eight feature vectors and the 3D coordinates of the point. These interpolated features from different resolution spaces are then concatenated to form the final encoding of the sample point. This encoded representation is first processed by the density MLP to predict the density, and the color MLP uses the density prediction results to estimate the color. The volume rendering process follows the same procedure as in the original NeRF.

\subsection{Computing-In-Memory Architecture}

Computing-in-Memory (CIM) has emerged as a promising solution to the memory bandwidth bottleneck by eliminating the need for data transfer between the computing unit and memory in the traditional von Neumann architecture, thereby reducing communication overhead~\cite{mutlu2022modern, sebastian2020memory,wang2024compass,liu2024paap}. The Resistive Random-Access-Memory~(ReRAM) is a two-terminal device with a metal oxide layer sandwiched between two metal electrode layers. 
For the ReRAM Single-Level Cell (SLC), its resistance can switch between the Low-Resistance State (LRS) and High-Resistance State (HRS), depending on the write voltage $V_{\mathrm{write}}$ applied across the electrodes~\cite{yuan2021forms}.
The ReRAM crossbar is known for its superior Vector-Matrix Multiplication (VMM) acceleration capability, computed via its in-situ current weighted summation operation~\cite{chi2016prime, ankit2019puma}. To compute $\bm{x}\cdot\bm{w}$ where $\bm{x},\bm{w} \in \{0,1\}^n$, elements of $\bm{x}$ and $\bm{w}$ are mapped as Digital-to-Analog Converter (DAC) outputs~($V_\mathrm{DAC}$) and ReRAM conductance~($G_\mathrm{LRS}, G_\mathrm{HRS}$) respectively. The computation output is obtained by measuring the output current of the source-line by an Analog-to-Digital Converter (ADC).

\makeatletter
\newcommand\figcaption{\def\@captype{figure}\caption}
\newcommand\tabcaption{\def\@captype{table}\caption}
\makeatother

\begin{figure}[tp]
    \setlength{\abovecaptionskip}{0pt}
    \setlength{\belowcaptionskip}{1pt}
    \centering
    \includegraphics[width=0.9\linewidth]{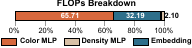}
    \caption{\small FLOPs breakdown.}
    \vspace{-0.3cm}
    \label{fig: flops breakdown}
\end{figure}

\section{Challenges and Motivation}
\label{sec:challenge}

\textbf{\emph{Challenge 1: Frequent \& Irregular Data Access on Large Embedding Tables.}}
Instant-NGP utilizes 16 embedding tables for multi-resolution encoding, each sized at $2^{19}$, resulting in a total embedding table size of approximately 60 MB. Although the time complexity of hash table-based encoding is $O(1)$, this approach remains memory-intensive due to the large size of the hash tables, which often exceeds the on-chip memory capacity of many edge devices. The hash-based indexing mechanism, combined with a high volume of sampling points, results in frequent and irregular memory accesses, leading to significant data transfer and performance inefficiencies. As illustrated in Figure \ref{fig: analysis}(b), each sampling point maps to multiple voxels in the resolution space, with each voxel containing 8 vertices that require feature lookups, resulting in extremely frequent data access. 

\textbf{Key Insight: }As a result, executing multi-resolution hash encoding naively would necessitate frequent off-chip memory accesses, imposing significant stress on the off-chip memory subsystem. Figure \ref{fig: access-analysis} illustrates address accesses for 1,500 consecutive sample points, highlighting that hash mapping generates accesses with poor spatial locality, which further exacerbates data handling overhead. Therefore, CIM is a promising solution to mitigate these data access challenges.

\begin{figure*}[tp]
    \setlength{\abovecaptionskip}{2pt}
    \setlength{\belowcaptionskip}{2pt}
    \centering
    \includegraphics[width=1\linewidth]{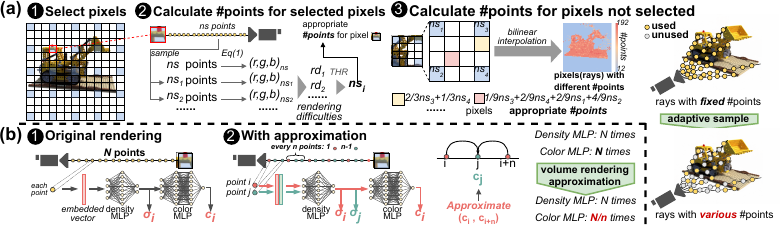}
    \caption{\small Rendering Optimization Overview. (a) Adaptive Sampling: Each pixel corresponds to a ray, making the terms ``pixel'' and ``ray'' equivalent in this context. (b) Volume Rendering Approximation.}
    \vspace{-0.2cm}
    \label{fig: framework}
\end{figure*}

\textbf{\emph{Challenge 2: High Energy Consumption of MLP Execution.}}
Although Instant-NGP reduces computational intensity by using embedded tables for 3D scene information storage, it still faces significant energy challenges. Each pixel in the image requires density and color predictions for hundreds of sampling points, resulting in tens of millions of MLP evaluations for a single rendered image. For instance, in NeRF, each ray involves $192$ samples, and to render an image at $800 \times 800$ resolution, the network must process over $100$ million inputs. While ReRAM-based CIM implementations can mitigate latency through efficient crossbar-based VMMs, the sheer volume of MLP computations results in substantial energy consumption, as shown in Figure~\ref{fig: flops breakdown}. Thus, a critical challenge is to reduce the number of MLP executions without compromising rendering quality to optimize energy efficiency.

\textbf{Key Insight: }We observe that each embedded feature of a sampled point must be processed by both a density MLP and a color MLP sequentially. Notably, the computational load of these two MLPs differs significantly: the density MLP accounts for only about 8\% of the total MLP FLOPs, while the color MLP accounts for 92\%. This disparity highlights a key opportunity for optimization: focusing on reducing the computational burden of the color MLP can lead to significant overall improvements in efficiency.

\textbf{\emph{Challenge 3: Data Access Conflicts in the Hash Table.}}
When rendering rays for neighboring pixels, the sample points that are closer to the camera often reside in the same or adjacent voxels. This situation can lead to simultaneous access requests to the same hash table when multiple sample points are processed in parallel---a scenario more common with low-resolution embedding tables. 

\textbf{Key Insight: }Despite CIM's potential to address \textbf{\emph{Challenge 1}} and \textbf{\emph{Challenge 3}}, effective resource utilization in the CIM architecture is challenging because multiple resolution embedding tables are stored within a single crossbar, with each row containing entries from different tables. This setup may result in simultaneous read operations from tables at different resolutions, requiring serial processing of these reads. Such serial access significantly slows down data access and reduces rendering efficiency. Therefore, an effective data mapping strategy that reduces the total number of accesses and minimizes conflict rates is crucial.

To address \textbf{\emph{Challenge 1}}, \textbf{\emph{2}}, and \textbf{\emph{3}} comprehensively, we propose Adaptive Sampling and Data Reuse (ASDR) solution with algorithmic innovations and CIM-based architecture. The core concept of ASDR is to dynamically reduce and reuse sample points, leveraging the inherent spatial locality in rendered images. This includes inter-ray locality (discussed in Section \ref{subsection: adaptive sample} and \ref{subsubsection: cache}), intra-ray locality (covered in Sections \ref{subsubsection: map} and \ref{subsubsection: cache}), and color-wise locality (explored in Section \ref{subsection: decouple}). For instance, neighboring sampled rays often share sampled points within the same voxel, and adjacent sample points along a single ray are also likely to be located within the same 3D voxel. This fine-grained locality presents new opportunities and motivates us to accelerate the rendering (inference) process of NeRF.

\section{ASDR Algorithm}

\subsection{Algorithm Overview}
To reduce data access and computational overhead in neural rendering, we optimize sampling at the algorithmic level by leveraging the inherent spatial locality present in rendered images. Our approach focuses on two key types of locality.
\textbf{Inter-ray Locality:} We propose an adaptive sampling scheme based on pixel rendering difficulty. This method dynamically adjusts the number of sampling points per pixel, optimizing both data access and computation.
\textbf{Color-wise Locality:} We introduce a volume rendering approximation scheme that decouples color and density predictions. By separating these tasks, we significantly reduce the computational demands on the MLPs.

The first method reduces the total number of sample points, thereby minimizing data access and computation needs. The second method decreases computational overhead by streamlining MLP operations. Together, these strategies enhance rendering efficiency without compromising image quality.

\subsection{Adaptive Sampling with Rendering Difficulty Awareness}
\label{subsection: adaptive sample}

\begin{figure}[tp]
    \setlength{\abovecaptionskip}{2pt}
    \setlength{\belowcaptionskip}{2pt}
    \centering
    \includegraphics[width=1\linewidth]{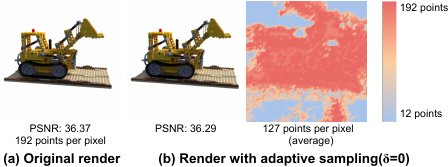}
    \caption{\small Sample optimization visualization. (a) Original rendering result for the LEGO scene. (b) Rendering result based on the adaptive sampling strategy, configured with $d=5$ and $\delta=0$. The \textcolor{red}{redder} pixels indicate a higher number of sample points required for rendering the corresponding ray, while \textcolor{blue}{bluer} pixels indicate fewer sample points needed.}
    \label{fig: sample visual}
    \vspace{-0.2cm}
\end{figure}

In NeRF models, such as Instant-NGP, the number of sampling points per pixel is typically fixed (e.g., 192 points are sampled along the ray corresponding to each pixel for rendering the LEGO scene). However, we observe that background pixels, which account for 40\% of the total pixels, can maintain rendering quality with significantly fewer samples (as low as 12 points). To address this inefficiency while maintaining real-time performance for online rendering, we propose an adaptive sampling scheme. The scheme first evaluates rendering difficulty by analyzing color variance in a subset of pixels, and then dynamically adjusts the number of sampling points for each pixel based on this evaluation.  This approach reduces unnecessary computations while preserving rendering quality, making it particularly suitable for online rendering scenarios where real-time adaptation is crucial.

As shown in Figure~\ref{fig: framework}(a), we begin by sampling a subset of pixels from the image for the first-stage rendering. For an image with a resolution of $D \times D$, we sample $\frac{D}{d} \times \frac{D}{d}$ pixels, where $d$ is the distance between neighboring pixels in both the horizontal and vertical directions. For each sampled pixel, we sample points along its corresponding ray according to the original fixed number of samples, denoted as $ns$. After predicting the color and density for all sampled points on a ray, we perform multiple volume renderings using different numbers of sampled points. Based on the differences among these rendering results, we determine a specific number of samples needed for each pixel.

Specifically, for a given ray with $ns$ sampled points, we perform volume rendering with varying numbers of points: $ns, ns_1, ns_2, \dots, ns_p$, where $ns_i$s and $p$ are preconfigured. This approach yields $p$ s-RGB color prediction results for each pixel, denoted as $(r,g,b){ns}, (r,g,b){ns_1}, \dots, (r,g,b)_{ns_p}$. Based on these rendering outcomes, we design a specific metric $rd_i$ to evaluate the rendering difficulty using $ns_i$ samples, as shown in Eq~(\ref{eq: rd}).
\begin{equation}
\label{eq: rd}
    rd_i = max (|r_{ns}-r_{ns_i}|, |g_{ns}-g_{ns_i}|, |b_{ns}-b_{ns_i}|)
\end{equation}
where $r_{ns_i}, g_{ns_i}, b_{ns_i}$ are the color values of $(r,g,b)_{ns_i}$.
The smaller the $rd_i$, the more similar the results rendered with $ns_i$ points are to those rendered with full $ns$ samples.
When $rd_i = 0$, it means that using $ns_i$ points can achieve lossless rendering for the pixel.

Based on the rendering difficulty $rd_i$s, we use a pre-set threshold $\delta$ to determine the number of sampling points for the pixel.
Specifically, we find the smallest $ns_i$ such that $rd_i \leq \delta$ and set the number of samples for that pixel to $ns_i$.
For the pixels that are not initially sampled, we calculate the number of sampling points by bilinear interpolation using the surrounding four sampled pixels. 
As shown in Figure~\ref{fig: sample visual}, we visualize the rendering results of the LEGO scene at different thresholds and the number of sampling points per pixel. In the visualization, blue indicates a low number of sampling points, while red indicates a high number. We observe that, compared to using a fixed 192 sample points per pixel, our adaptive sampling scheme achieves almost the same rendering quality with an average of 120 sample points per pixel. Our scheme's ability to adaptively select fewer samples for background pixels and areas with simple textures is key to achieving lossless optimization.

\begin{figure}[tp]
    \setlength{\abovecaptionskip}{2pt}
    \setlength{\belowcaptionskip}{2pt}
    \centering
    \includegraphics[width=1\linewidth]{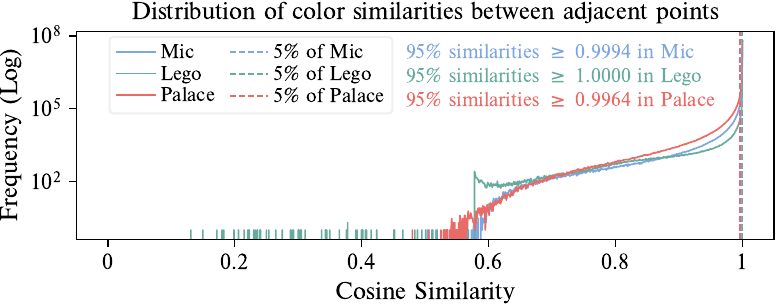}
    \caption{\small Distribution Analysis of Cosine Similarities Between Adjacent Sampled Points Along Rays. $95\%$ of cosine similarities are close to $1$, indicating that adjacent points along the rays have highly similar colors. This high similarity underscores the significant spatial locality of point colors (\textbf{color-wise locality}).} 
    \label{fig: jieou insight}
    \vspace{-0.1cm}
\end{figure}

\begin{figure}[bp]
    \vspace{-0.4cm}
    \setlength{\abovecaptionskip}{2pt}
    \setlength{\belowcaptionskip}{0pt}
    \centering
    \includegraphics[width=1\linewidth]{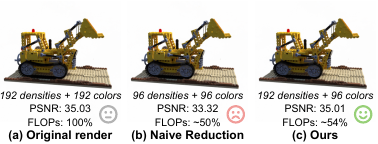}
    \caption{\small Visualization of Volume Rendering Approximation. (a) Original LEGO scene rendering. (b) Rendering result with naive sampling point reduction by half. (c) Rendering result with approximation optimization based on color-wise locality.}
    \label{fig: approx vis}
\end{figure}

\begin{figure*}[tp]
    \centering
    \setlength{\abovecaptionskip}{2pt}
    \setlength{\belowcaptionskip}{2pt}
    \includegraphics[width=1\linewidth]{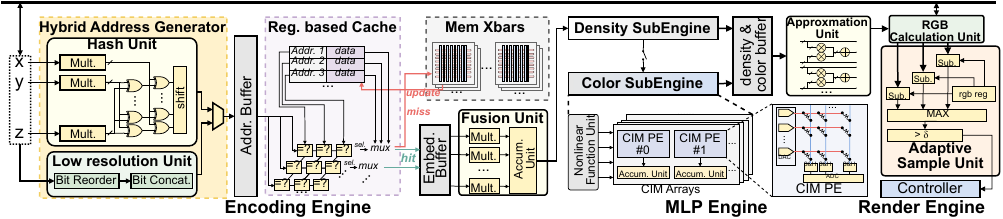}
    \caption{\small Overview of the ASDR Architecture.}
    \label{fig: arch}
    % \vspace{-0.1cm}
\end{figure*}

\subsection{Rendering Approximation based on Color-Wise Locality}
\label{subsection: decouple}

Adaptive sampling allows for simultaneous optimization of data access and MLP computation by reducing the number of sampling points. In addition to this, we further aim to minimize the significant amount of MLP computation by exploiting color-wise locality.
As illustrated in Figure~\ref{fig: framework}(b), the MLP structure used for neural rendering is generally divided into two parts: the density network and the color network. The density network processes the embedded features of the sampled points to produce a density result and a 15-dimensional feature vector, which is then fed into the color network for color computation. As discussed in Section~\ref{sec:challenge}, the color network in the MLP phase is responsible for the majority of the computational overhead compared to the density network. Therefore, minimizing the execution of the color network is crucial for improving overall performance.

We analyze the distribution of cosine similarity between the RGB colors of adjacent sampled points along the rays, as shown in Figure~\ref{fig: jieou insight}. Our analysis reveals that more than 95\% of the similarities are very close to 1, illustrating the high color-wise spatial locality of the sampled points.
Based on this insight, we propose an approximation-based color density decoupling method that leverages neighboring sample points. For a ray with $N_c$ sampled points, we divide these points into $\frac{N_c}{n}$ groups, each containing $n$ points. Within the $i$-th group, we compute the color for only one point, $c_{n(i-1)+1}$, which denotes the first sampling point of the $i$-th group, using the color network. For the remaining points, we estimate their colors through linear interpolation, utilizing the distance between sampled points and the computed color information from points $c_{(i-1)n+1}$ and $c_{in+1}$. During volume rendering, the interpolated results serve as approximate colors for the sampled points that do not undergo color MLP computation.

This volume rendering optimization significantly reduces MLP computation with marginal quality loss, as illustrated in Figure~\ref{fig: approx vis}. In the configuration where $n=2$, we maintain nearly the same PSNR\footnote{PSNR (Peak Signal-to-Noise Ratio) measures rendering quality, with higher values indicating superior results.} while achieving a 46\% reduction in computation. Figure~\ref{fig: approx vis}(b) further shows that our approximation strategy achieves $\approx 1.7$ PSNR improvement compared to merely halving ($\geq 50\%$) the number of sampling points.

\section{ASDR Architecture}

\subsection{Architecture Overview}
\label{sec: 5.1}
To support the proposed algorithm-level optimizations, we design an architecture that incorporates several optimizations. Our architecture, named ASDR, is tailored to improve data layout, address mapping, and locality utilization, with a strong emphasis on data reuse.

Figure~\ref{fig: arch} provides an overview of the ASDR architecture, which consists of three main components: an encoding engine, an MLP engine, and a volume rendering engine.
\begin{enumerate}
    \item \textbf{Encoding Engine}: Optimizes embedding table lookups via enhanced storage/address mapping and a register-based cache, exploiting inter-ray and intra-ray locality to minimize redundancy.
    \item \textbf{MLP Engine}: Uses CIM-based crossbar memory with MAC functionality (CIM PEs) for efficient color/density predictions, performing MVM in-memory to eliminate parameter data movement.
    \item \textbf{Volume Rendering Engine}: Implements adaptive sampling and approximation units for algorithm-level optimizations, enabling efficient sampling and color-density decoupling to reduce computations.
\end{enumerate}

\subsection{Encoding Engine}

\begin{figure*}[tp]
    \centering
        \begin{minipage}[c][0.12\textheight][c]{0.68\textwidth}
        \centering
        \includegraphics[width=1\linewidth]{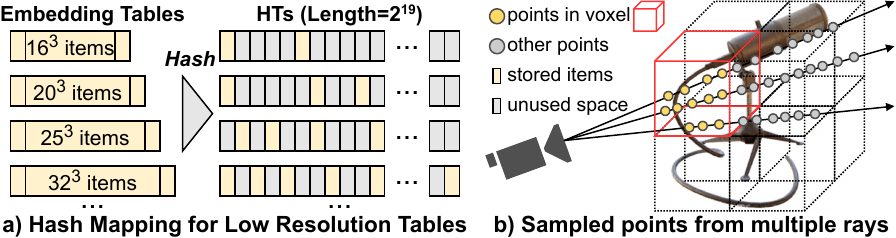}
        % % \vspace{-1cm}
        \figcaption{\small Mapping and Rendering Analysis for Architecture Optimization.}
        \label{fig: arch-moti}
    \end{minipage}
    \begin{minipage}[c][0.17\textheight][t]{0.28\textwidth}
        \centering
        \includegraphics[width=1\linewidth]{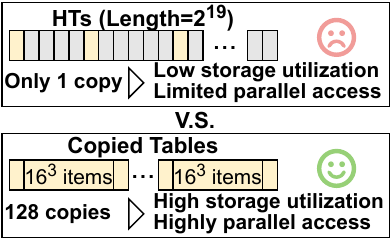}
        \figcaption{\small Copy for low resolution tables.}
        \label{fig: arch-tablecopy}
    \end{minipage}%
    % \vspace{-0.1cm}
\end{figure*}

\begin{figure}[htbp]
    \setlength{\abovecaptionskip}{2pt}
    \setlength{\belowcaptionskip}{2pt}
    \centering
    \includegraphics[width=1\linewidth]{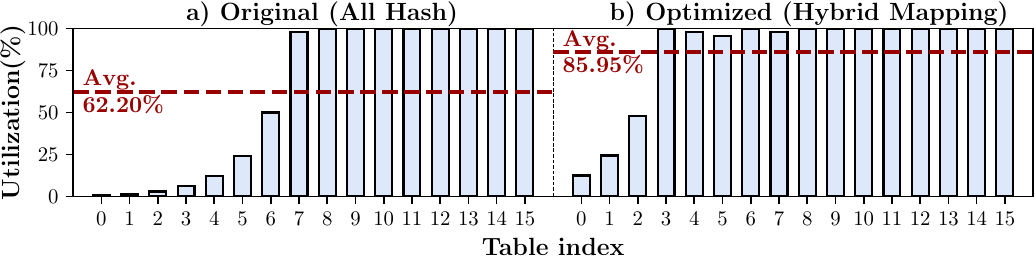}
    \caption{\small Storage Utilization before/after Optimization.}
    \label{fig: profiling-storage}
    \vspace{-0.2cm}
\end{figure}

The encoding engine comprises a hybrid address generator, a register-based cache, memory crossbars (Mem Xbars) for storing embedding tables, and a fusion unit. The address generator creates addresses to index the embedded features based on the sample point information sent from the bus. A register-based cache utilizes locality to reduce read conflicts in the Mem Xbars, while the fusion unit merges features of different resolutions corresponding to the sampling points to complete the feature encoding.

\subsubsection{Hybrid Address Generator for Fewer Conflicts}
\label{subsubsection: map}
NeRF models typically require embedding tables of varying resolutions to model scenes at different granularities. In the original model, high-resolution tables are compressed using hash indexing to reduce storage overhead. This means some 3D space points share the same table entries due to hash conflicts. While this approach effectively compresses storage for high-resolution tables, it results in unnecessary storage waste for low-resolution tables.

Figures \ref{fig: arch-moti}(a) and \ref{fig: profiling-storage}(a) illustrate that naively storing all embedding tables in memory crossbars according to hash address mapping can lead to approximately 38\% resource waste. This observation inspired us to explore whether the extra storage space could be utilized for access optimization. Additionally, low-resolution tables often face severe access conflicts. As depicted in Figure \ref{fig: arch-moti}(b), several consecutive sample points on a ray can be indexed in the low-resolution space using the embedded features of eight vertices corresponding to the same 3D cube. Consequently, each of these eight embedded features may be indexed multiple times. In contrast, high-resolution spaces tend to have sampling points corresponding to different 3D cubes, resulting in a lower probability of Mem Xbar read conflicts due to hash-generated indexes.

\begin{figure}[tp]
    \setlength{\abovecaptionskip}{2pt}
    \setlength{\belowcaptionskip}{2pt}
    \centering
    \includegraphics[width=1\linewidth]{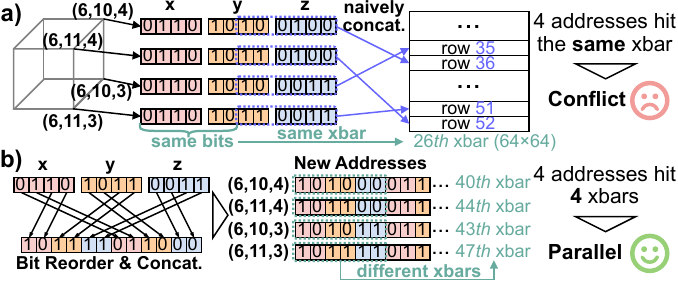}
    \caption{\small Address generation without hashing.}
    \label{fig: arch-address}
    \vspace{-0.2cm}
\end{figure}

Dynamically determining and de-duplicating duplicate indexes in hardware is inefficient. Therefore, we utilize the observed storage headroom by adopting a de-hashed indexing approach for low-resolution tables and replicating multiple copies for storage. The most straightforward way to de-hash an index is to concatenate the $(x,y,z)$ coordinates of the vertices to construct the address, as shown in Figure \ref{fig: arch-address}(a). However, this approach can lead to intra-cube conflicts. For instance, if the $z$-coordinate is used as the last few bits of the address, the high bits of the addresses for the four vertices of a cube will be identical, leading to read conflicts since the corresponding embedded entries are stored on the same Mem Xbar. 

To mitigate this issue, we select the low bits of the vertices' $(x,y,z)$ coordinates to form the high bits of the address, as depicted in Figure \ref{fig: arch-address}(b). Specifically, we generate new addresses through bit reordering and concatenation, ensuring that vertices are stored on different Mem Xbars and thereby enhancing parallel access. Since low-resolution embedded tables are stored without hash mapping, there are unused spaces. We duplicate these embedded tables to utilize the remaining space, refining address generation by appending the copy IDs of the embedded tables as the high bits of the address.
Figure~\ref{fig: arch-tablecopy} illustrates an example of copying a low-resolution table with $16^3$ items. Before copying, only $\frac{16^3}{2^{19}} = \frac{1}{128}$ of the storage is utilized. After copying, the storage is fully utilized, enabling highly parallel access for efficient encoding.

For high-resolution tables, we maintain the original hash mapping for storage and address generation using the hash unit shown in Figure \ref{fig: arch}. As Figure \ref{fig: profiling-storage}(b) demonstrates, the optimized storage space utilization with hybrid mapping reaches 85.95\%, which is nearly 25\% higher than the original mapping scheme.

\subsubsection{Register-based Cache for Higher Locality Utilization}
\label{subsubsection: cache}

In the low-resolution 3D grid space, two neighboring pixels on the imaging plane, corresponding to two rays, are highly likely to pass through the same 3D voxel, resulting in the same embedding index query. This occurs because the embedding of a sampling point at any position within a 3D voxel is computed based on trilinear interpolation of the embeddings of the eight vertices of that voxel. Additionally, neighboring sample points along a single ray are also likely to be distributed within the same 3D voxel. To capitalize on this observation, we analyzed the distribution properties of sampled points during rendering and identified two spatial localities (i.e., inter-ray and intra-ray locality) that can be exploited.

\begin{figure}[t!]
    \setlength{\abovecaptionskip}{2pt}
    \setlength{\belowcaptionskip}{2pt}
    \centering
    \includegraphics[width=0.95\linewidth]{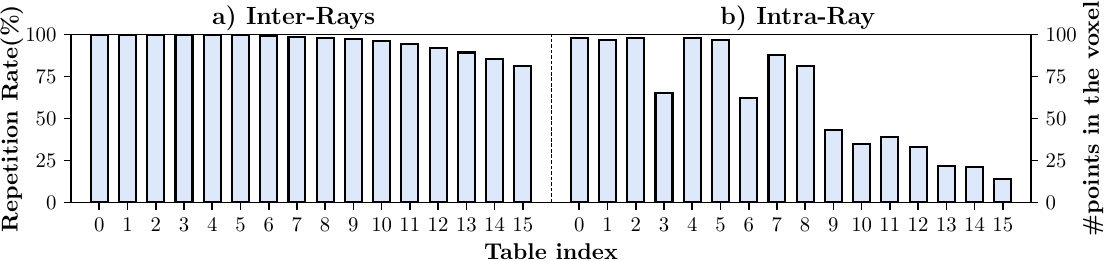}
    \caption{\small Profiling of Point Repetition Rates in Sampling from Rays. (a) \textbf{Inter-Ray}: Average repetition rate of sample points from neighboring rays, highlighting the spatial locality between rays. (b) \textbf{Intra-Ray}: Distribution of sample points within the voxel containing the highest number of points for a selected ray, illustrating the concentration of sampled points within the same voxel.}
    \label{fig: profiling-sample}
    \vspace{-0.2cm}
\end{figure}

First, we examined the repetition rate of sample points between neighboring rays (pixels), as shown in Figure~\ref{fig: profiling-sample}(a). For 12 out of the 16 resolution spaces, there is a repetition rate of 90\% or more for sample points between neighboring rays. Even in the highest resolution space, the repetition rate is over 70\%, indicating extremely high spatial locality of sample points between rays. It's important to note that repetition here does not imply that the coordinates of the sampled points are exactly the same, but rather that the sampled points correspond to the same voxels.

Additionally, Figure~\ref{fig: profiling-sample}(b) shows the number of sample points within the voxel containing the most points from a single ray, illustrating the repetition of points along the ray.
Across all resolution spaces, each ray consists of 192 sampled points. For example, in the lowest resolution space (hash table 1), 98 out of the 192 sampled points are located within the same voxel. Even in the highest resolution space, 21 sample points are located within the same voxel, demonstrating the high spatial locality of sample points within a ray.

Such a heightened repetition rate of sampled points indicates that both inter-ray and intra-ray sampling points exhibit high spatial locality, often residing in the same voxel grid. This means that the same embeddings are indexed for feature computation, as these embeddings correspond to the eight vertices of the voxel cubes. By exploiting this locality, we can reduce the frequency of accesses to the embedding table. To achieve this, we designed a register-based cache that bypasses repeated accesses to identical embeddings in an embedding table.

For each resolution-space embedding table, we utilize a register to cache the most recently accessed table entries. During each clock cycle, a batch of addresses is generated. These addresses are derived from the vertex coordinates by identifying the voxel cubes in which the sampled points are located, and they are further refined using a hybrid address generator. For these addresses, we implement all-to-all comparison circuits to enable parallel queries. Each address is compared against all table entry addresses in the cache. If a match is found, the system bypasses accessing the memory crossbars and directly outputs the cached embedded features to the fusion unit.
If no match is found, the address is sent to the Mem Xbars for a table lookup operation, and the retrieved embedding, along with its corresponding address, is written to the cache as a new table entry. The cache employs a Least Recently Used (LRU) replacement strategy. Cache sizes vary for different resolution embedded tables based on the locality of sampling points.

\subsubsection{Fusion Unit}

The fusion unit is responsible for computing the features of each sampled point based on the embeddings of the eight voxel vertices found in each resolution space. The calculation of these features requires trilinear interpolation, which is performed using the eight embedding vectors and coordinate information.
To achieve this, the fusion unit employs several multipliers and accumulators to perform the weighted accumulation of the eight embedding vectors. Once the interpolation is complete, the features obtained at different resolutions are concatenated to achieve feature fusion, resulting in a unified representation for each sampled point.

\subsection{MLP Engine}

The MLP engine is composed of two sub-engines: the density sub-engine and the color sub-engine. Each sub-engine comprises nonlinear function units and CIM arrays, where each sub-engine is responsible for implementing activation functions and normalization operations. The CIM array is composed of multiple CIM PEs and associated accumulation units to efficiently perform vector-matrix multiplication. The CIM PEs are crossbar memory arrays equipped with multiply-accumulate (MAC) functionality. To accommodate the density-color decoupling optimization proposed in the previous section, the pathway from the density sub-engine to the color sub-engine is designed to be skippable. For sample points processed using the decoupling scheme, the color computation is bypassed once the density computation is completed and stored in the cache. In contrast, sample points not utilizing the decoupling approximation proceed to the color sub-engine for color prediction, with the results subsequently cached.

\subsection{Volume Rendering Engine}

The volume rendering engine is composed of three main components: the approximation unit, the RGB computation unit, and the adaptive sampling unit. 
Approximation Unit implements the density-color decoupling approximation scheme. It uses a series of multipliers and adders to perform linear interpolation based on the computed color information of the sampled points, allowing it to approximate the color of uncomputed sampled points.
RGB Computation Unit is responsible for calculating the RGB values of pixels corresponding to rays. It uses the color and density information of all the sampling points along the rays, including those supplemented by the approximation unit.
Adaptive Sampling Unit supports adaptive sampling optimization by determining the necessary number of sampling points for each sampled pixel. It employs subtractors and comparators to calculate the required number of sampling points as per Eq.~(\ref{eq: rd}). The computed results are then sent to the controller to allocate the appropriate number of sampling points for all pixels.

\subsection{Dataflow}

ASDR performs image rendering in two distinct phases:
\textbf{Phase I: Initial Computation for Adaptive Sampling.} In the first phase, ASDR calculates the required number of sampling points for each pixel. This is done by processing several sampled pixels from the bus, using the adaptive sampling strategy to determine the optimal sampling points. During this phase, all architectural components are engaged. The computed number of sampling points for each pixel is sent to the controller, setting the stage for the rendering phase.
\textbf{Phase II: Full Image Rendering.} In the second phase, ASDR renders the entire image based on the sampling points predicted in the first phase. This phase focuses on generating the RGB values for each pixel. Unlike the initial computation, the results from this phase are sent back to the bus as the final output, without routing through the adaptive sampling unit.
During the process, sample points are handled as follows: 
(1) \textbf{Address Generation}: Sample points are directed to the hybrid address generator, which allocates them to the appropriate address unit based on their 3D resolution space. Points in high-resolution spaces are sent to the hash unit, while those in low-resolution spaces are sent to the low-resolution unit. The address generation supports parallel processing, with addresses buffered in an address buffer for simultaneous indexing of the embedded features.
(2) \textbf{Embedding Lookup}: Given the variability in access times (cache miss \textit{v.s.} cache hit), indexed data is stored in the embed buffer while awaiting for all necessary embedded features to be retrieved. The retrieved data is then fed into the fusion unit to compute the sampled point features.
(3) \textbf{MLP Processing}: The MLP engine processes the encoded features for color and density prediction. For efficiency, some sample points may bypass the color MLP sub-engine. Results are buffered in the density \& color buffer.
(4) \textbf{Color Approximation}: After processing all sample points on a ray, the approximation unit estimates colors for points that lack predicted values, using interpolation based on the computed colors of other points on the same ray. The refined color information is then passed to the RGB computation unit, which completes the color calculation for each pixel.

\begin{table}[tp]
\setlength{\abovecaptionskip}{3pt}
\setlength{\belowcaptionskip}{2pt}
\caption{\small Dataset Statics.}
\resizebox{1\linewidth}{!}{
\addtolength{\tabcolsep}{-3pt} 
\renewcommand\arraystretch{1.1}
\begin{tabular}{lccc}
\Xhline{1.5px}
Dataset & Scene & Resolution & Type \\ \Xhline{1.2px}
Synthetic-NeRF~\cite{nerf} & Mic, Hotdog, Ship, Chair, Ficus, Lego & 800$\times$800 & \multirow{2}{*}{Synthetic} \\ \cline{1-3}
Synthetic-NSVF~\cite{liu2020neural} & Palace & 800$\times$800 &  \\ \hline
BlendedMVS~\cite{knapitsch2017tanks} & Fountain & 768$\times$576 & \multirow{3}{*}{\begin{tabular}[c]{@{}c@{}}Real\\ World\end{tabular}} \\ \cline{1-3}
Tanks\&Temples~\cite{yao2020blendedmvs} & Family & 1920$\times$1080 &  \\ \cline{1-3}
Instant-NGP~\cite{instngp} & Fox & 1080$\times$1920 &  \\ \Xhline{1.5px}
\end{tabular}
}
\label{tab: dataset}
% % \vspace{-0.5cm}
\end{table}

\begin{table}[tp]
\setlength{\abovecaptionskip}{3pt}
\setlength{\belowcaptionskip}{2pt}
\caption{\small Configuration of ASDR(-Server/-Edge).}
\resizebox{1\columnwidth}{!}{
\addtolength{\tabcolsep}{-2pt} 
\renewcommand\arraystretch{1.1}
\begin{tabular}{clccc}
\Xhline{1.5px} % \hline
\multicolumn{2}{c}{Component} & \begin{tabular}[c]{@{}c@{}}Area\\ ($mm^2$)\end{tabular} & \begin{tabular}[c]{@{}c@{}}Power\\ ($mW$)\end{tabular} & Config \\ \Xhline{1.2px}
\multirow{4}{*}{\begin{tabular}[c]{@{}c@{}}Encoding\\ Engine\end{tabular}} & Address Generator & 0.013 / 0.003 & 8.04 / 2.01 & 64 / 16 \\ 
 & Reg-based Cache & 0.007 / 0.002 & 2.66 / 0.67 & 128 / 32 \\
 & Mem Xbars & 5.03 / 1.26 & 5.33 / 1.33 & 64MB / 2MB \\ 
 & Fusion Unit & 0.220 / 0.055 & 107.99 / 27.00 & 32 / 8 \\ \hline
\multirow{2}{*}{\begin{tabular}[c]{@{}c@{}}MLP\\ Engine\end{tabular}} & Density SubEngine & 3.44 / 0.86 & 28.44 / 7.11 & 4 / 1 \\ 
 & Color SubEngine & 5.76 / 1.44 & 47.30 / 11.82 & 4 / 1 \\ \hline 
\multirow{3}{*}{\begin{tabular}[c]{@{}c@{}}Render \\ Engine\end{tabular}} & Approximation Unit & 0.118 / 0.029 & 52.21 / 13.05 & 16 / 4 \\ 
 & RGB Unit & 0.013 / 0.003 & 5.40 / 1.35 & 8 / 2 \\ 
 & Adaptive Sample Unit & 0.0007 / 0.0002 & 0.27 / 0.07 & 8 / 2 \\ \hline
Buffers &  & 0.27 / 0.06 & 79 / 19.55 & 256KB / 64KB \\ \hline
Total & ASDR-Server/-Edge & 15.09 / 3.77 & 5.77W / 1.44W & - \\ \Xhline{1.5px}
\end{tabular}
}
\label{tab: config}
\end{table}

\section{Evaluation}
\label{sec: exp}

\subsection{Evaluation Methodology}
\label{sec: 6.1}

\textbf{Datasets and Metrics.} To evaluate the rendering quality and performance improvement achieved by our proposed ASDR design, we conduct experiments using five commonly used datasets, as detailed in Table~\ref{tab: dataset}. Seven scenes are from synthetic datasets and three are from the real world, each characterized by high resolution and rendering complexity. We evaluate image quality using the Peak Signal-to-Noise Ratio (PSNR) metric, with higher PSNR values indicating superior rendering quality.
We also use SSIM (Structural Similarity Index) and LPIPS (Learned Perceptual Image Patch Similarity) for more objective measurements.
Higher SSIM values indicate greater structural similarity to the ground truth.
Lower LPIPS values indicate better perceptual similarity to the ground truth.

\begin{figure*}[tp]
    \setlength{\abovecaptionskip}{2pt}
    \setlength{\belowcaptionskip}{2pt}
    \centering
    \includegraphics[width=1\linewidth]{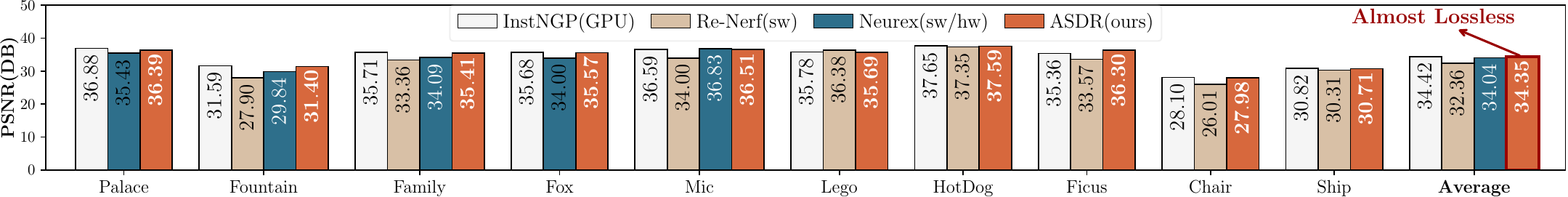}
    \caption{\small Rendering Quality Comparison.}
    \label{exp-fig: accuracy}
    \vspace{-0.2cm}
\end{figure*}

\textbf{Baselines.} To evaluate the performance of the ASDR design, we compare it against three types of computing platforms: edge devices, consumer-level GPUs, and specialized NeRF accelerators.
1). Edge Device: We use the NVIDIA Jetson Xavier NX~\cite{xavier} as a representative edge device, providing insights into performance on portable hardware.
2). Consumer-level GPU: The NVIDIA RTX 3070\footnote{\url{https://www.nvidia.com/en-us/geforce/graphics-cards/30-series/}.} is selected to represent high-end consumer-level rendering acceleration hardware. We utilize Instant-NGP's CUDA implementation to perform NeRF inference on the GPU, recording execution times.
3). NeRF Accelerator: We compare our design with NeuRex~\cite{neurex}, a state-of-the-art NeRF accelerator. 
For a fair comparison with the GPUs, we scale the number of computing cores to ensure the same area budget across platforms. The execution times on the GPUs are scaled based on the ratio of the number of cores in the GPUs to those in the ASDR architecture. To compare with NeuRex, we construct a cycle-accurate simulator that accounts for NeuRex’s performance losses, such as grid cache misses and hardware underutilization.

\textbf{Implementation.}
We implement the proposed ASDR algorithm based on
the open-source code\footnote{\url{https://github.com/NVlabs/instant-ngp}.} of the most representative NeRF model, Instant-NGP~\cite{instngp}, running on the PyTorch framework. To evaluate the performance of the ASDR architecture, we developed a cycle-level simulator that mimics the hardware behavior of the encoding engine, MLP engine, and rendering engine. 
For digital circuit implementation, we used Verilog and synthesized the RTL with Synopsys Design Compiler using the TSMC 28nm standard library at 1GHz to evaluate the area and power consumption of the designed engines. We employ CACTI~\cite{balasubramonian2017cacti} to estimate the energy and area of each on-chip buffer, considering their width, size, and number of reads/writes. For CIM components, we simulated the memory crossbars and the MLP engine using NeuroSim~\cite{chen2018neurosim}. Each CIM crossbar is configured to a size of $64 \times 64$, with a 5-bit ADC precision in the MLP engine. We integrated the RTL synthesis results and simulated CIM results into the simulator.

We evaluated two variants of ASDR: ASDR-Edge and ASDR-Server. ASDR-Edge is designed under strict area and power constraints, which is typical for edge platforms. In contrast, ASDR-Server is a scaled-up architecture intended for high-end computing platforms. Detailed ASDR configurations are provided in Table~\ref{tab: config}.

\begin{table}[tp]
\setlength{\abovecaptionskip}{2pt}
\setlength{\belowcaptionskip}{2pt}
\caption{\small More rendering quality metrics comparison of ASDR against Instant-NGP.}
\resizebox{1\linewidth}{!}{
\addtolength{\tabcolsep}{-2pt} 
\begin{tabular}{ccccccccc}
\Xhline{1.5px}
\multirow{2}{*}{\textbf{Metrics}} & \multirow{2}{*}{\textbf{Models}} & \multicolumn{7}{c}{\textbf{Scenes}} \\ \cline{3-9} 
 &  & \textbf{Lego} & \textbf{Ship} & \textbf{Hotdog} & \textbf{Chair} & \textbf{Mic} & \textbf{Ficus} & \textbf{Average} \\ \Xhline{1.5px}
\multirow{2}{*}{\textbf{SSIM}} & Instant-NGP & 0.986 & 0.938 & 0.985 & 0.985 & 0.986 & 0.982 & 0.977 \\ \cline{2-9} 
 & ASDR & 0.985 & 0.931 & 0.983 & 0.983 & 0.985 & 0.982 & 0.975 \\ \hline
\multirow{2}{*}{\textbf{LPIPS}} & Instant-NGP & 0.037 & 0.137 & 0.049 & 0.055 & 0.035 & 0.061 & 0.062 \\ \cline{2-9} 
 & ASDR & 0.039 & 0.143 & 0.052 & 0.057 & 0.035 & 0.062 & 0.064 \\ \Xhline{1.5px}
\end{tabular}
}
\label{tab: other metric}
\vspace{-0.3cm}
\end{table}

\begin{figure}[tp]
    \setlength{\abovecaptionskip}{2pt}
    \setlength{\belowcaptionskip}{2pt}
    \centering
    \includegraphics[width=1\linewidth]{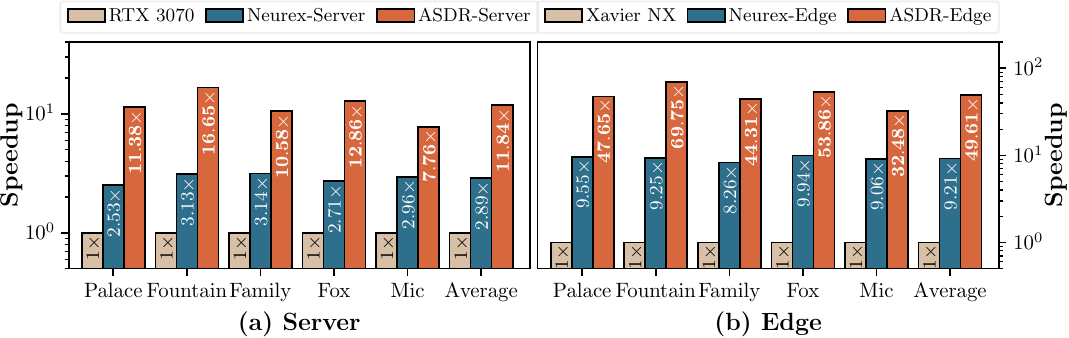}
    \caption{\small Speedup of ASDR compared to GPU and NeuRex in (a) Server and (b) Edge configurations.}
    \label{exp-fig: speedup}
    \vspace{-0.3cm}
\end{figure}

\subsection{Algorithm Evaluation}
\label{sec: exp acc more metric}

\textbf{Rendering Quality.}
Figure~\ref{exp-fig: accuracy} compares the rendering quality of our ASDR design with the original Instant-NGP model and two other optimization techniques: Re-NeRF and Neurex. The evaluation results demonstrate that ASDR achieves nearly lossless performance compared to the original model. On average, our design only reduced the PSNR by 0.07 compared to Instant-NGP, while Re-NeRF and Neurex showed larger decreases in PSNR, with reductions of 2.06 and 0.38, respectively.
In the HotDog scene, ASDR exhibited only a 0.06 decrease in PSNR, indicating that our adaptive sampling scheme effectively identifies pixel rendering difficulties. It also mitigates quality degradation typically caused by reducing the number of sampled points. Additionally, the rendering approximation scheme based on color-wise locality successfully approximates the color of sampled points.

Interestingly, ASDR achieved even better rendering performance than the original model in the Ficus scene. This suggests that our adaptive sampling scheme can filter out noise by customizing the number of sampling points for each pixel, enhancing the overall rendering quality.

We further compare ASDR's rendering quality against Instant-NGP using additional metrics: SSIM and LPIPS. As shown in Table~\ref{tab: other metric}, ASDR achieves rendering quality comparable to Instant-NGP, with an average difference of $0.002$ in both SSIM and LPIPS.

\begin{figure}[tp]
    \setlength{\abovecaptionskip}{2pt}
    \setlength{\belowcaptionskip}{2pt}
    \centering
    \includegraphics[width=1\linewidth]{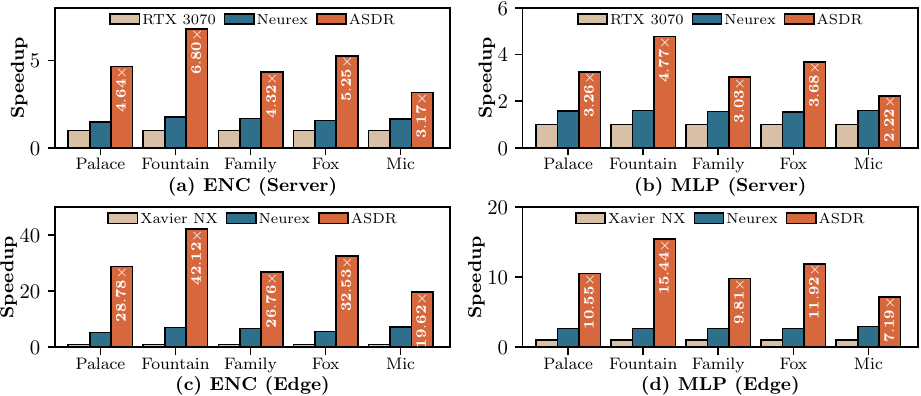}
    \caption{\small Speedup on hash encoding and MLP phase.}
    \label{exp-fig: speedup breakdown}
    \vspace{-0.2cm}
\end{figure}

\begin{figure}[tp]
% % \vspace{-0.3cm}
    \setlength{\abovecaptionskip}{2pt}
    \setlength{\belowcaptionskip}{2pt}
    \centering
    \includegraphics[width=1\linewidth]{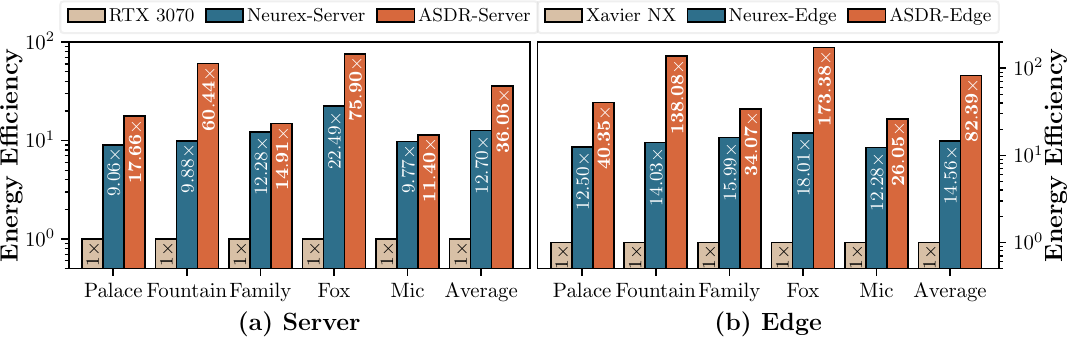}
    \caption{\small Energy Efficiency of ASDR compared to GPU and NeuRex in (a) Server and (b) Edge configurations.}
    \label{exp-fig: energy}
    \vspace{-0.2cm}
\end{figure}

\subsection{Overall Performance}

\textbf{Speedup.} 
Figure~\ref{exp-fig: speedup} illustrates the significant speedup achieved by ASDR compared to GPU and accelerator baselines. On average, ASDR-Server provides a $11.84\times$ speedup over the RTX 3070 GPU and $4.11\times$ over Neurex-Server, while ASDR-Edge offers a speedup of $49.61\times$ over the Xavier NX and $5.38\times$ over Neurex-Edge. These improvements are due to ASDR’s efficient use of spatial locality at different levels, which reduces computational complexity, maximizes data reuse, and minimizes data access conflicts. 

We further analyze the speedup by breaking it down into two phases: encoding and MLP execution. As shown in Figure~\ref{exp-fig: speedup breakdown}, ASDR-Server achieves a $3.90\times$ speedup in encoding and a $2.77\times$ speedup in MLP execution compared to the baselines. ASDR-Edge achieves a $17.37\times$ speedup in encoding and a $7.52\times$ speedup in MLP execution. Adaptive sampling improves speedup in both phases by reducing the number of sampled points, which lowers data access requirements. In addition, rendering approximation boosts speedup by cutting the number of color MLP executions by at least half. In the encoding stage, our data mapping and reuse optimizations significantly reduce read conflicts. Most existing solutions do not address these issues, which is why ASDR achieves greater speedup in encoding compared to the MLP phase.

\begin{figure}[tp]
    \setlength{\abovecaptionskip}{2pt}
    \setlength{\belowcaptionskip}{2pt}
    \centering
    \includegraphics[width=1\linewidth]{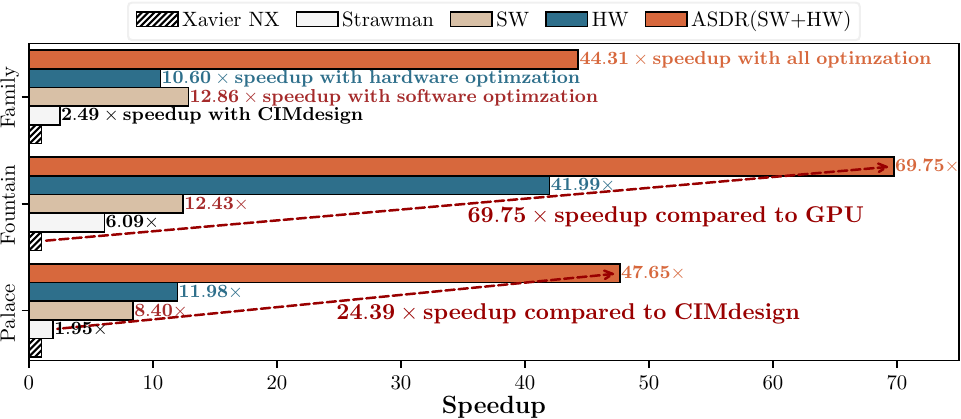}
    \caption{\small Detailed analysis of contributions.}
    \label{exp-fig: ablation}
    \vspace{-0.2cm}
\end{figure}

\begin{figure}[tp]
    \setlength{\abovecaptionskip}{2pt}
    \setlength{\belowcaptionskip}{2pt}
    \centering
    \includegraphics[width=1\linewidth]{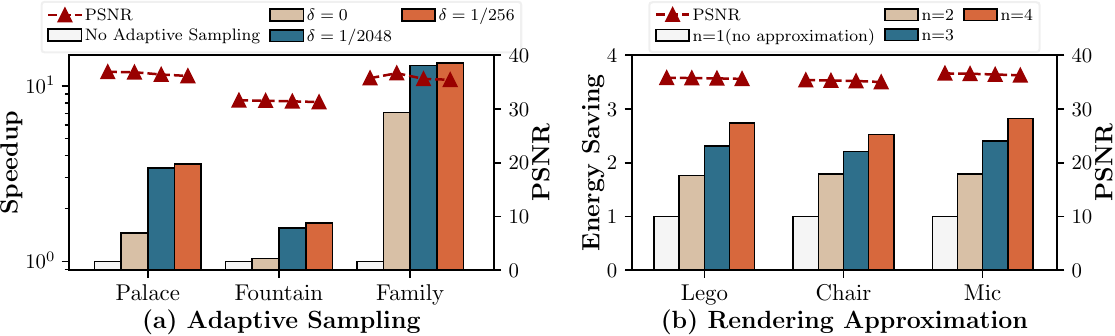}
    \caption{\small Analysis of the threshold $\delta$ for adaptive sampling and rendering approximation.}
    \label{exp-fig: threshold+jieou}
    \vspace{-0.2cm}
\end{figure}

\textbf{Energy Efficiency.}
Figure~\ref{exp-fig: energy} compares the energy efficiency of ASDR with baseline systems. On average, ASDR achieves $59.22\times$ and $4.08\times$ higher energy efficiency compared to GPUs and accelerators. This improvement is primarily due to our optimization techniques, which reduce read conflicts and minimize stalls in the rendering pipeline. Additionally, ASDR's design incurs minimal area and power costs, preserving the high energy efficiency of CIM arrays.

\subsection{Ablation Study}

Figure~\ref{exp-fig: ablation} shows the results of our ablation studies. The strawman design, a basic CIM approach, achieves only a $3.51\times$ speedup compared to the Xavier NX, indicating it does not fully exploit CIM-based architecture. Single hardware or software optimizations achieve $11.23\times$ and $21.52\times$ speedup, respectively. In contrast, our ASDR design, which combines both hardware and software optimizations, achieves an impressive $53.90\times$ speedup. For the fountain scene, ASDR provides a $69.75\times$ speedup over the GPU, and for the palace scene, a $24.39\times$ speedup over the strawman design. 

\subsection{Design Space Exploration}
\label{sec: DSE}

We explore the design space for adaptive sampling, rendering optimization, and data reuse.

\textbf{Adaptive Sampling Threshold:} Figure~\ref{exp-fig: threshold+jieou}(a) shows that a threshold of $\delta=1/2048$ achieves a $6.02\times$ speedup with less than $0.3$ PSNR loss. Higher thresholds provide diminishing returns.

\textbf{Rendering Approximation Group Size:} Figure~\ref{exp-fig: threshold+jieou}(b) indicates that with a group size of $n=4$, ASDR can save about $2.7\times$ in energy with less than $0.3$ PSNR loss.

\begin{figure}[tp]
    \setlength{\abovecaptionskip}{2pt}
    \setlength{\belowcaptionskip}{2pt}
    \centering
    \includegraphics[width=1\linewidth]{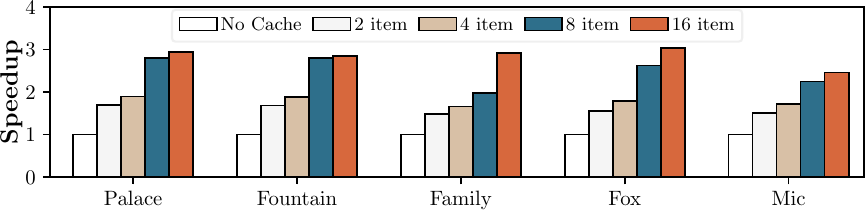}
    \caption{\small Analysis of the cache size.}
    \label{exp-fig: cache}
    \vspace{-0.2cm}
\end{figure}

\textbf{Cache Size:} Figure~\ref{exp-fig: cache} demonstrates that using a cache with $8$ items per embedding table provides a $2.49\times$ speedup compared to no cache. This low-cost cache leverages observed data locality effectively.

\begin{figure}[tp]
    \setlength{\abovecaptionskip}{2pt}
    \setlength{\belowcaptionskip}{2pt}
    \centering
    \includegraphics[width=1\linewidth]{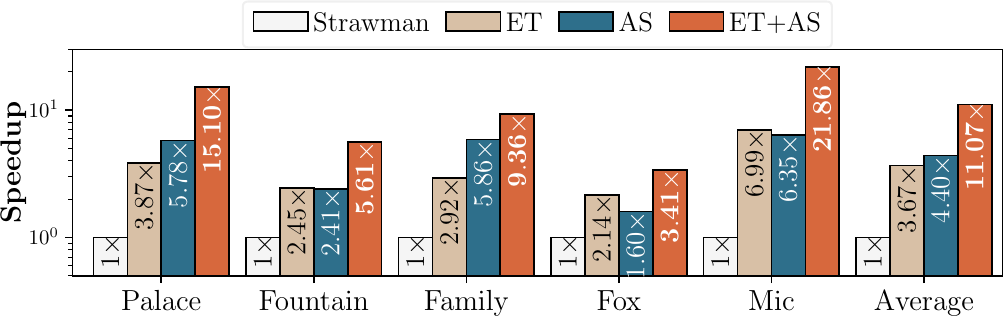}
    \caption{\small Performance analysis of ASDR with early termination and adaptive sampling. Strawman: ASDR without early termination or adaptive sampling. ET: ASDR with early termination only. AS: ASDR with adaptive sampling only.}
    \label{exp-fig: revision early teminate}
    % \vspace{-0.4cm}
\end{figure}

\subsection{Combined Acceleration with Early Termination}
\label{sec: exp early temination}

The proposed adaptive sampling is orthogonal to the widely-used early termination technique~\cite{instngp}, which terminates pixel rendering when accumulated opacity exceeds $1$ to avoid redundant computation. We evaluate ASDR's performance when integrated with early termination, as shown in Figure~\ref{exp-fig: revision early teminate}.
ASDR with adaptive sampling alone achieves an average $4.4\times$ speedup over the strawman design, while ASDR with only early termination achieves $3.67\times$ speedup. Combining both techniques enables ASDR to achieve an average $11.07\times$ speedup, demonstrating the significance of the proposed adaptive rendering. Importantly, since early termination does not alter the volume rendering algorithm, the rendering quality remains unaffected.

\begin{figure}[tp]
    \setlength{\abovecaptionskip}{2pt}
    \setlength{\belowcaptionskip}{2pt}
    \centering
    \includegraphics[width=1\linewidth]{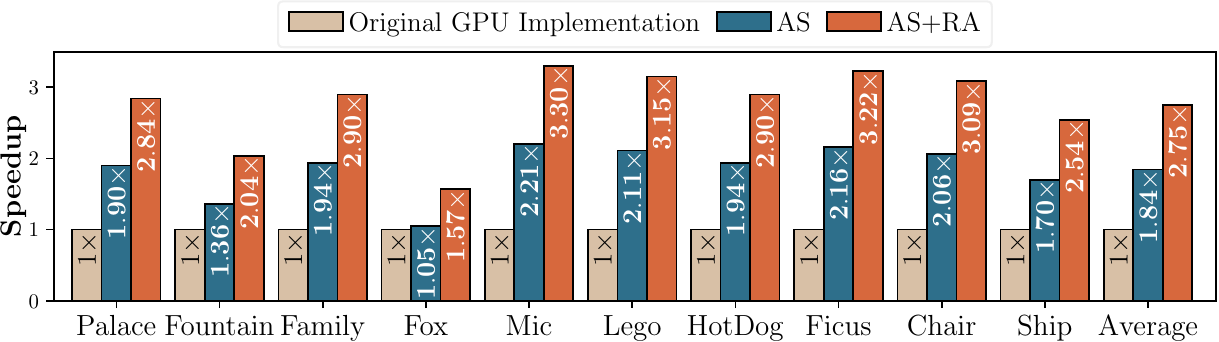}
    \caption{\small GPU performance of software-level optimizations without hardware acceleration. AS: GPU with adaptive sampling. AS+RA: GPU with adaptive sampling and rendering approximation (color-wise locality).}
    \label{exp-fig: revision no hardware}
    \vspace{-0.3cm}
\end{figure}

\subsection{Performance without Hardware Acceleration}
\label{sec: exp no hardware gpu}

To evaluate the effectiveness of our software-level optimizations without hardware acceleration, we implement adaptive sampling and rendering approximation in CUDA on RTX 3070. As shown in Figure~\ref{exp-fig: revision no hardware}, adaptive sampling alone achieves a $1.84\times$ speedup over the original GPU implementation. When combined with rendering approximation, the speedup increases to an average of $2.75\times$. These results highlight the significant performance gains achievable through our software-level optimizations, demonstrating their potential for accelerating rendering without relying on specialized hardware.

\subsection{Performance on TensoRF}
\label{sec: exp tensorf}

We further evaluate ASDR on TensoRF~\cite{tensorf}, a widely used NeRF model. Table~\ref{tab: tensorf} shows that ASDR preserves nearly lossless rendering quality. Figure~\ref{exp-fig: tensorf speedup} presents the speedup results, where our software-only GPU implementation achieves $1.27\times$ acceleration, primarily due to adaptive sampling reducing the number of points. The ASDR architecture achieves an average speedup of up to $29.98\times$, highlighting effective data access optimization.

\subsection{Performance Across Hardware Configurations}
\label{sec: exp no reram dependency}

We evaluate the performance of ASDR across different hardware configurations to demonstrate its adaptability beyond ReRAM-based implementations. ASDR (SA) combines SRAM-based memory for encoding with a systolic array for MLP; ASDR (SRAM) uses SRAM-based memory and CIM macros for MLP; ASDR (ReRAM) is the native ReRAM-based implementation. SRAM components are modeled using NeuroSim~\cite{chen2018neurosim}, and the systolic array follows Eyeriss~\cite{8686088}. Experimental results (Figures~\ref{exp-fig: revision no reram speedup} and~\ref{exp-fig: revision no reram energy}) show ASDR (SA) achieves $8.90\times$/$3.08\times$ speedup and $18.22\times$/$1.43\times$ energy efficiency over RTX 3070/Neurex-Server. ASDR (SRAM) outperforms GPU by $9.53\times$ and Neurex-server by $3.3\times$, with a $1.31\times$ energy efficiency advantage over ASDR (SA) due to SRAM-based CIM macros. These results validate our optimizations' generalization across hardware architectures.

\section{Related Works}
\label{sec: 7}

\noindent\textbf{Neural Rendering Acceleration}
Recent advancements in neural rendering have led to the development of various accelerators aimed at improving performance. For instance, ICARUS~\cite{rao2022icarus} and RT-NeRF~\cite{RT-NeRF} design specialized architectures to optimize the computationally intensive MLP execution in vanilla neural rendering models~\cite{nerf}. 
Gen-NeRF~\cite{Gen-NeRF} enhances TensoRF~\cite{tensorf}, a parametric encoding-based model, through a coarse-to-fine sampling strategy. Inst-3D~\cite{Instant-3D} and Combricon-R~\cite{Cambricon-R} improve training efficiency for Instant-NGP~\cite{instngp} by leveraging embedding grid decomposition and ray-based execution. NeuRex~\cite{neurex} analyzes the inference process of Instant-NGP and proposes a subgrid-based approach for hardware-efficient encoding.

\begin{table}[tp]
\setlength{\abovecaptionskip}{2pt}
\setlength{\belowcaptionskip}{2pt}
\caption{\small Rendering Quality of ASDR on TensoRF~\cite{tensorf}.}
\resizebox{1\linewidth}{!}{
\addtolength{\tabcolsep}{-3pt} 
\begin{tabular}{ccccccccccccc}
\Xhline{1.5px}
\multirow{2}{*}{\textbf{Metrics}} & \multirow{2}{*}{\textbf{Models}} & \multicolumn{11}{c}{\textbf{Scenes}} \\ \cline{3-13} 
 &  & \textbf{Palace} & \textbf{Fountain} & \textbf{Family} & \textbf{Fox} & \textbf{Mic} & \textbf{Lego} & \textbf{Hotdog} & \textbf{Ficus} & \textbf{Chair} & \textbf{Ship} & \textbf{AVG} \\ \Xhline{1.5px}
\multirow{2}{*}{\textbf{PSNR}} & \textbf{TensoRF} & 35.27 & 31.03 & 33.41 & 34.75 & 35.09 & 35.22 & 36.48 & 34.13 & 35.01 & 30.32 & 34.07 \\ \cline{2-13} 
 & \textbf{ASDR} & 35.25 & 30.94 & 33.17 & 34.51 & 35.02 & 35.1 & 36.24 & 34.12 & 34.82 & 30.11 & 33.93 \\ \hline
\multirow{2}{*}{\textbf{SSIM}} & \textbf{TensoRF} & 0.977 & 0.910 & 0.904 & 0.908 & 0.988 & 0.983 & 0.98 & 0.987 & 0.981 & 0.904 & 0.952 \\ \cline{2-13} 
 & \textbf{ASDR} & 0.977 & 0.9 & 0.899 & 0.903 & 0.985 & 0.979 & 0.971 & 0.985 & 0.974 & 0.897 & 0.947 \\ \hline
\multirow{2}{*}{\textbf{LPIPS}} & \textbf{TensoRF} & 0.021 & 0.132 & 0.101 & 0.216 & 0.017 & 0.022 & 0.038 & 0.024 & 0.025 & 0.135 & 0.073 \\ \cline{2-13} 
 & \textbf{ASDR} & 0.021 & 0.137 & 0.110 & 0.223 & 0.017 & 0.024 & 0.039 & 0.024 & 0.028 & 0.136 & 0.076 \\ \Xhline{1.5px}
\end{tabular}
}
\label{tab: tensorf}
\vspace{-0.2cm}
\end{table}

\begin{figure}[tp]
    \setlength{\abovecaptionskip}{2pt}
    \setlength{\belowcaptionskip}{2pt}
    \centering
    \includegraphics[width=1\linewidth]{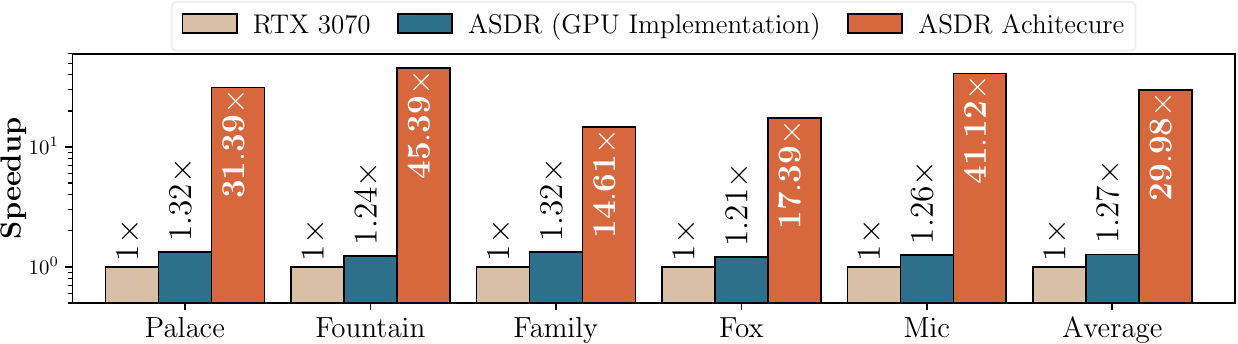}
    \caption{\small Performance of ASDR on TensoRF~\cite{tensorf}.}
    \label{exp-fig: tensorf speedup}
    \vspace{-0.2cm}
\end{figure}

\begin{figure*}[tp]
    \setlength{\abovecaptionskip}{2pt}
    \setlength{\belowcaptionskip}{2pt}
    \centering
    \includegraphics[width=1\linewidth]{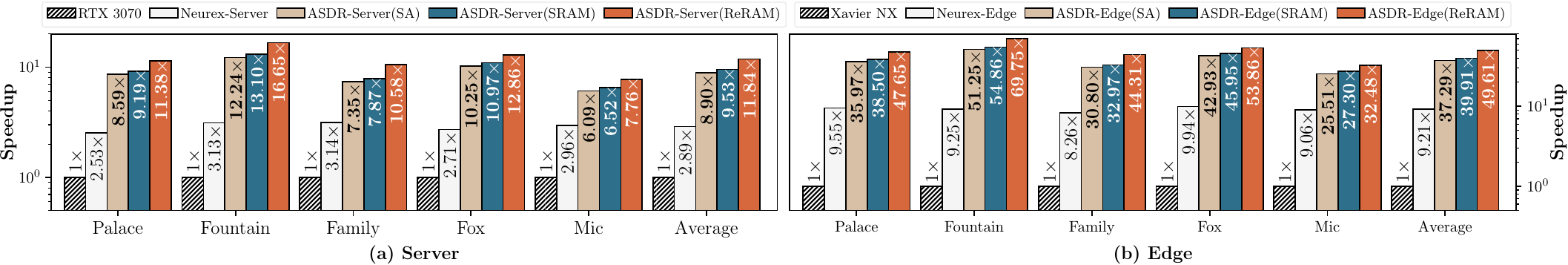}
    \caption{\small The speedup of ASDR across different hardware configurations. ASDR(SA): SRAM-based encoding engine with systolic array for MLP; ASDR(SRAM): SRAM-based encoding engine with SRAM CIM macros for MLP; ASDR(ReRAM): ReRAM-based encoding engine with ReRAM CIM macros for MLP (native implementation).}
    \label{exp-fig: revision no reram speedup}
    \vspace{-0.1cm}
\end{figure*}

\begin{figure*}[tp]
    \setlength{\abovecaptionskip}{2pt}
    \setlength{\belowcaptionskip}{2pt}
    \centering
    \includegraphics[width=1\linewidth]{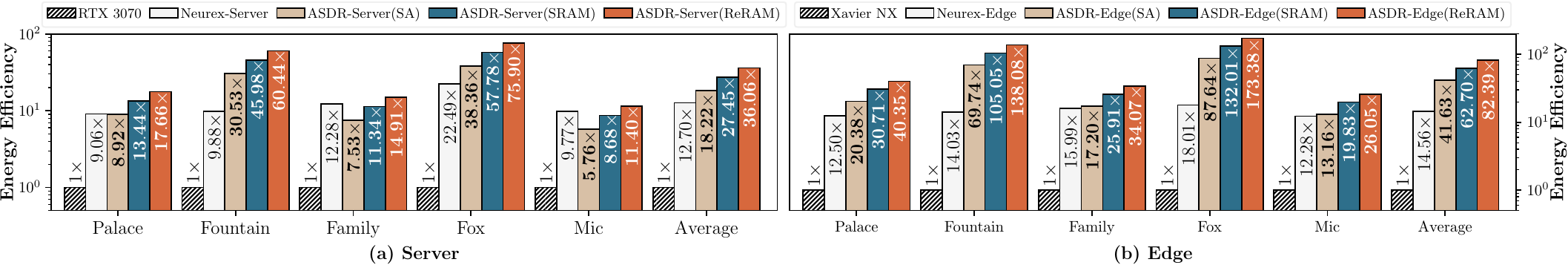}
    \caption{\small Energy efficiency of ASDR without a dependence on ReRAM-based hardware. Notations of ASDR(SA), ASDR(SRAM), and ASDR(ReRAM) are the same as Figure~\ref{exp-fig: revision no reram speedup}.}
    \label{exp-fig: revision no reram energy}
    \vspace{-0.1cm}
\end{figure*}

Cicero~\cite{feng2024cicero} is a recent accelerator for Instant-NGP-like models, proposing memory-centric rendering to reduce DRAM accesses. However, its requirement to store intermediate results on-chip poses significant power challenges. In contrast, ASDR avoids pipeline modifications and optimizes irregular access via hybrid address generation, focusing on single-frame sampling-point-level similarity. These methods are complementary, and adaptive sampling in ASDR could further enhance Cicero by reducing sampling points.

\noindent\textbf{CIM-based Acceleration}
CIM-based architecture is increasingly utilized in neural network acceleration research to address the data transfer limitations of traditional von Neumann systems. ReHY~\cite{jin2021rehy} introduces a hybrid CIM architecture combining digital and analog ReRAM for accelerating CNNs. ReTransformer~\cite{yang2020retransformer} uses ReRAM crossbars to achieve high energy efficiency in transformer acceleration. TransPIM~\cite{zhou2022transpim} develops a data flow optimized for CIM-friendly transformer inference.
AttAcc~\cite{10476442} enhances the efficiency of batched inference in large language models through execution optimization.

\section{Discussion}
\label{sec: discuss}

This section discusses two key questions: (1) the applicability of our optimizations to more NeRF models, and (2) a comparison to 3D Gaussian Splatting (3DGS)~\cite{kerbl20233d}.

\subsection{Applicability to More NeRF Models}
\label{sec: discuss more nerf models}
We focus on Instant-NGP as a representative NeRF model due to its widespread adoption, state-of-the-art performance, and efficient use of multi-resolution hash grids, as highlighted in Table~\ref{tab: more models}. However, our optimizations are designed to generalize across NeRF variants. For instance: \textbf{1) DirectVoxGO}~\cite{sun2022direct}: Uses multi-resolution 3D grids similar to Instant-NGP, making our optimizations directly applicable.
\textbf{2) TensorRF}~\cite{tensorf}: Models features with tensors across multiple planes, indexed as embedding tables, requiring minimal modifications for our approach.
To demonstrate this, we evaluate ASDR on TensorRF in Section~\ref{sec: exp tensorf}, achieving significant speedup without compromising rendering quality. This underscores the generalizability of our design to other NeRF models.

\subsection{Comparison to 3D Gaussian Splatting}

3D Gaussian Splatting~\cite{kerbl20233d} addresses NeRF's efficiency bottlenecks using 3D Gaussian primitives for real-time rendering. However, it lacks neural network support, leading to lower rendering quality with significant input view deviations, and requires storing millions of Gaussian primitives, making it unsuitable for edge scenarios. 
Recently, a ton of works have been proposed to enhance the performance of 3DGS. 
GScore~\cite{gscore} introduces a more refined intersection test to reduce redundancy caused by Gaussian replication and designs an architecture with hierarchical sorting units to accelerate rendering. 
MetaSapiens~\cite{MetaSapiens} proposes an efficiency-aware pruning technique to optimize rendering speed and a foveated rendering (FR) method, leveraging the low sensitivity of human peripheral vision to dynamically adjust rendering quality and speed. 
GBU~\cite{GBU} designs a Gaussian Blending Unit that enhances data reuse and identifies unimportant Gaussian primitives. 
VR-pipe~\cite{vr-pipe} proposes a native hardware support for early termination and Uni-Render~\cite{uni-render} designs a unified accelerator adapted for various NeRF models.
Moreover, since 3DGS performs volume rendering like Instant-NGP, our adaptive sampling technique could be adapted as ``adaptive Gaussian sampling'', optimizing the number of Gaussian primitives per pixel or tile for efficient rendering. We plan to explore this in future work.

\begin{table}[tp]
\setlength{\abovecaptionskip}{2pt}
\setlength{\belowcaptionskip}{2pt}
\caption{\small Comparison with more NeRF Models.
}
\resizebox{1\linewidth}{!}{
\begin{tabular}{ccc}
\Xhline{1.5px}
\textbf{NeRF Models} & \textbf{Feature Modeling} & \textbf{Density/Color Comp.} \\ \Xhline{1.5px}
\textbf{DirectVoxGO~\cite{sun2022direct}} & multi-resolution 3D grids & interpolation + MLP \\ \hline
\textbf{TensoRF~\cite{tensorf}} & 2D grids (decomposed from 3D grids) & interpolation + MLP \\ \hline
\textbf{Instant-NGP~\cite{instngp}} & multi-resolution 3D grids + Hash & interpolation + MLP \\ \Xhline{1.5px}
\end{tabular}
}
\label{tab: more models}
\vspace{-0.1cm}
\end{table}

\section{Conclusion}

This paper presents ASDR, an algorithm-hardware codesign framework to facilitate an efficient rendering process of NeRF. We introduce a dynamic adaptive sampling scheme and a color-density decoupling strategy to reduce data access and computational overhead. At the architectural level, we propose a CIM-based architecture for efficient data mapping and reuse, which minimizes data access conflicts and improves rendering throughput. Experiments show that ASDR achieves significant speedups with trivial image quality loss.

\begin{acks}
We would like to thank the anonymous ASPLOS reviewers for their constructive feedback and suggestions.
We would also like to express our gratitude to Professor Yuhao Zhu from Rochester University who served as our shepherd.
We express our gratitude to Professor Yu Feng from Shanghai Jiao Tong University.
Our research benefited from discussion with him about his work Cicero~\cite{feng2024cicero}.

This work was partially supported by the National Key Research and Development Program of China (2024YFE0204300), National Natural Science Foundation of China (Grant No.62402311), and Natural Science Foundation of Shanghai (Grant No.24ZR1433700). 
Li Jiang and Fangxin Liu are the corresponding authors.
\end{acks}

\bibliographystyle{ACM-Reference-Format}
\bibliography{sample-base}

\end{document}